\documentclass[twocolumn]{pasj01}
\usepackage{mathpazo}
\usepackage[T1]{fontenc}

\begin{document} 
\Received{14-May-2018}
\Accepted{17-Jul-2018}

\title{A high dust emissivity index $\beta$ for a CO-faint galaxy in a filamentary Ly$\alpha$ nebula at $z=3.1$}
\author{Yuta Kato\altaffilmark{1,2}}
\author{Yuichi Matsuda\altaffilmark{1,3}}
\author{Daisuke Iono\altaffilmark{1,3}}
\author{Bunyo Hatsukade\altaffilmark{4}}
\author{Hideki Umehata\altaffilmark{4,5}}
\author{Kotaro Kohno\altaffilmark{4,6}}
\author{David M. Alexander\altaffilmark{7}}
\author{Yiping Ao\altaffilmark{1,8}}
\author{Scott C. Chapman\altaffilmark{9}}
\author{Matthew Hayes\altaffilmark{10}}
\author{Mariko Kubo\altaffilmark{1}}
\author{Bret D. Lehmer\altaffilmark{11}}
\author{Matthew A. Malkan\altaffilmark{12}}
\author{Tomonari Michiyama\altaffilmark{1,3}}
\author{Tohru Nagao\altaffilmark{13}}
\author{Tomoki Saito\altaffilmark{14}}
\author{Ichi Tanaka\altaffilmark{15}}
\author{Yoshiaki Taniguchi\altaffilmark{16}}
\altaffiltext{1}{National Astronomical Observatory of Japan, 2-21-1 Osawa, Mitaka, Tokyo, 181-8588, Japan}
\altaffiltext{2}{Department of Astronomy, Graduate School of Science, The University of Tokyo, 7-3-1 Hongo, Bunkyo-ku, Tokyo 133-0033, Japan}
\altaffiltext{3}{Department of Astronomy, School of Science, The Graduate University for Advanced Studies (SOKENDAI), Osawa, Mitaka, Tokyo, 181-8588, Japan}
\altaffiltext{4}{Institute of Astronomy, Graduate School of Science, The University of Tokyo, 2-21-1 Osawa, Mitaka, Tokyo 181-0015, Japan}
\altaffiltext{5}{RIKEN Cluster for Pioneering Research, 2-1 Hirosawa, Wako-shi, Saitama 351-0198, Japan}
\altaffiltext{6}{Research Center for the Early Universe, Graduate School of Science, The University of Tokyo, 7-3-1 Hongo, Bunkyo, Tokyo 113-0033}
\altaffiltext{7}{Centre for Extragalactic Astronomy, Department of Physics, Durham University, South Road, Durham, DH1 3LE, UK}
\altaffiltext{8}{Purple Mountain Observatory, Chinese Academy of Sciences, Nanjing 210034, China}
\altaffiltext{9}{Department of Physics and Atmospheric Science, Dalhousie University, Halifax, NS, B3H 4R2, Canada}
\altaffiltext{10}{Department of Astronomy and Oskar Klein Centre for Cosmoparticle Physics, Stockholm University, AlbaNova University Centre, SE-106 91 Stockholm, Sweden}
\altaffiltext{11}{Department of Physics, University of Arkansas, 226 Physics Building, 825 West Dickson, Fayetteville, AR 72701, USA}
\altaffiltext{12}{Department of Physics and Astronomy, University of California, Los Angeles, CA 90095, USA}
\altaffiltext{13}{Research Center for Space and Cosmic Evolution, Ehime University, 2-5 Bunkyo-cho, Matsuyama, Ehime 790-8577}
\altaffiltext{14}{Nishi-Harima Astronomical Observatory, Centre for Astronomy, University of Hyogo, 407-2 Nichigaichi, Sayo-cho, Sayo, Hyogo 679-5313, Japan}
\altaffiltext{15}{Subaru Telescope, National Astronomical Observatory of Japan, 650 North A’ohoku Place, Hilo, HI 96720, USA}
\altaffiltext{16}{The Open University of Japan, 2-11, Wakaba, Mihama-ku, Chiba, 261-8586}
\email{kato.yu@nao.ac.jp, yuta.astrophysics@gmail.com}



\maketitle

\begin{abstract}
We present CO\,$J=4-3$ line and 3\,mm dust continuum observations of a 100\,kpc-scale filamentary Ly$\alpha$ nebula (SSA22 LAB18) at $z=3.1$ using the Atacama Large Millimeter/submillimeter Array (ALMA). 
We detected the CO\,$J=4-3$ line at a systemic $z_{\rm{CO}}=3.093\pm0.001$ at 11\,$\sigma$ from one of the ALMA continuum sources associated with the Ly$\alpha$ filament.
We estimated the CO\,$J=4-3$ luminosity of $L'_{\rm{CO(4-3)}}=(2.3\ \pm\ 0.2)\times10^9$\,K\,km\,s$^{-1}$\,pc$^2$ for this CO source, which is one order of magnitude smaller than those of typical $z>1$ dusty star-forming galaxies (DSFGs) of similar far-infrared luminosity $L_{\rm{IR}}\sim10^{12}\,L_{\rm{\odot}}$.
We derived a molecular gas mass of $M_{\rm{gas}} = (4.4^{+0.9}_{-0.6}) \times 10^9\, M_{\rm{\odot}}$ and a star-formation rate of SFR$= 270 \pm 160\,M_{\rm{\odot}}\,\rm{yr}^{-1}$.
We also estimated a gas depletion time of $\tau_{\rm{dep}}=17\pm10\,\rm{Myr}$, being shorter than those of typical    DSFGs.
It is suggested that this source is in a transition phase from DSFG to a gas-poor, early-type galaxy.
From ALMA to \textit{Herschel} multi-band dust continuum observations, we measured a dust emissivity index $\beta=2.3\pm0.2$, which is similar to those of local gas-poor, early-type galaxies.
Such a high $\beta$ can be reproduced by specific chemical compositions for interstellar dust at the submillimeter wavelengths from recent laboratory experiments.
ALMA CO and multi-band dust continuum observations can constrain the evolutionary stage of high-redshift galaxies through $\tau_{\rm{dep}}$ and $\beta$, and thus we can investigate dust chemical compositions even in the early Universe.
\end{abstract}

\begin{figure*}
\begin{center}
\includegraphics[scale=0.3]{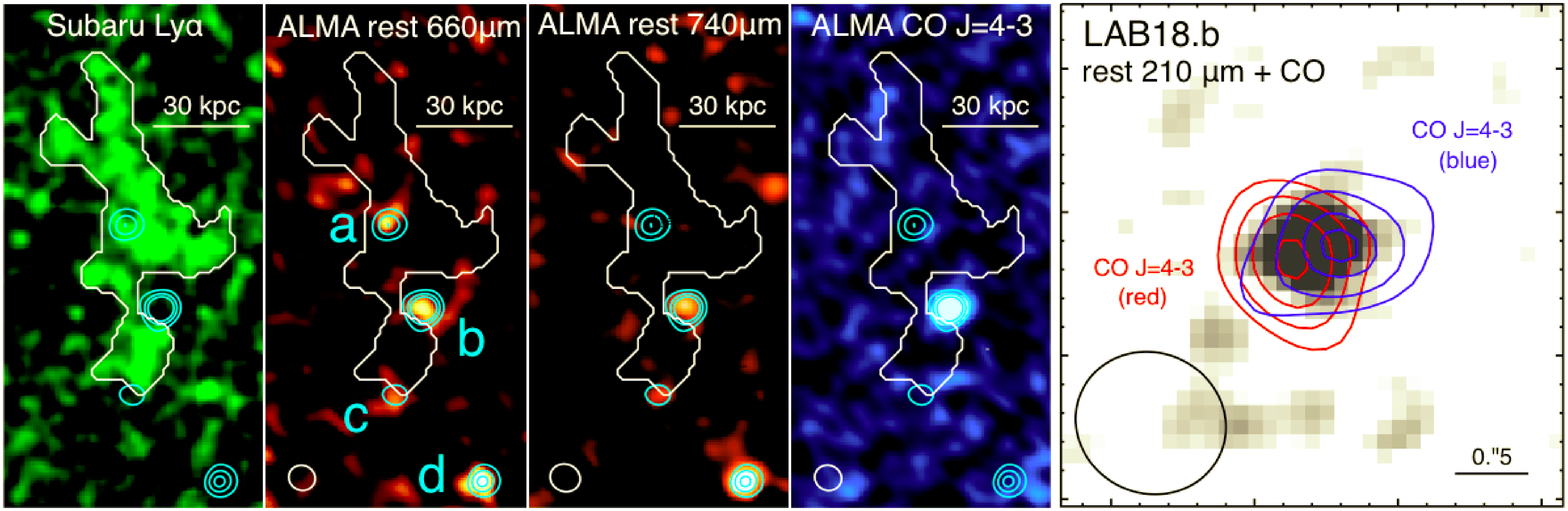}
\includegraphics[scale=0.14]{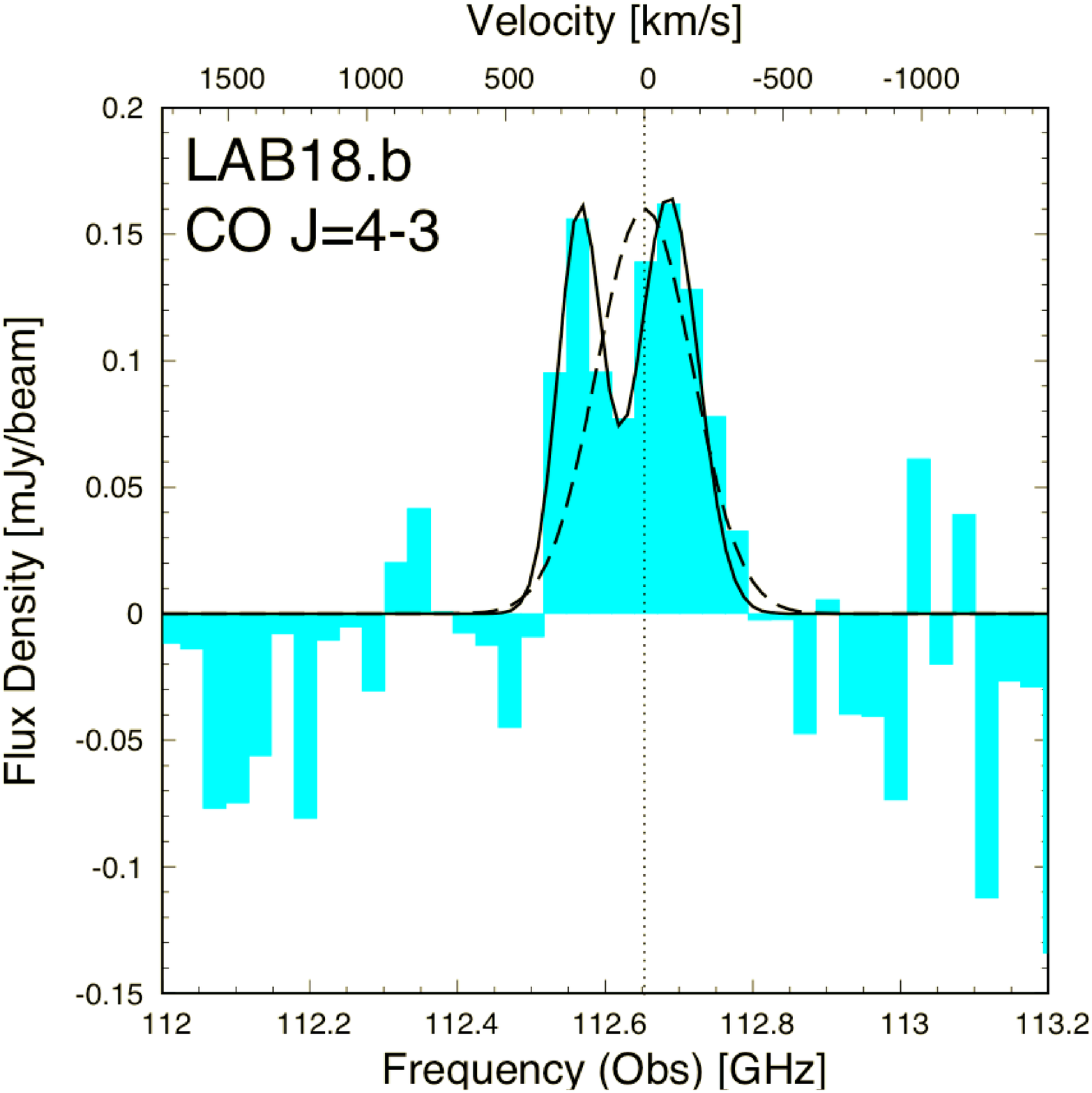}
\end{center}
\caption{\textit{(left)}. Thumbnail images of the 100\,kpc-scale filamentary Ly$\alpha$ nebula (SSA22 LAB18) at $z=3.1$. The image size is $10''\times20''$ ($\sim$80$\times$160 kpc$^2$).
From left to right, we show the Subaru/Suprime-Cam Ly$\alpha$ \citep{matsuda04}, ALMA rest-frame 660\,$\mu$m, 740\,$\mu$m continuum and ALMA CO\,$J=4-3$ (This work).
White contours show the Ly$\alpha$ emission with a surface brightness of $1.4\times10^{-18}$ erg s$^{-1}$ cm$^{-2}$ arcsec$^{-2}$.
Cyan contours show the ALMA rest-frame 210\,$\mu$m continuum (LAB18.a,\,b\,,c\,,d) possibly associated with the filamentary Ly$\alpha$ nebula (\cite{ao18}, Matsuda et al.\,in\,prep).
\textit{(middle)}. The red and blue components of CO\,$J=4-3$ lines for LAB18.b, which are shown as red and blue contours overlaid the ALMA rest-frame 210\,$\mu$m continuum. The Red and blue contour levels are 3, 4, 5, 6\,$\sigma$ and 3, 4, 5, 5.5\,$\sigma$, respectively. The spatial offset between the two components is $0\farcs4$ or $\sim$\,3\,kpc. The image size is $3\farcs5\times3\farcs5$ ($\sim$30$\times$30 kpc$^2$). The ellipses at the bottom-left corner represent the beam FWHM for the ALMA observations for left and right panels. \textit{(right)}. The CO\,$J=4-3$ spectrum for LAB18.b with a velocity resolution of 80\,km s$^{-1}$.
The best-fit single and double Gaussian profiles are overlaid as dashed and solid curves, respectively.}
\label{fig1}
\end{figure*}

\section{Introduction}
Ly$\alpha$ blobs (LABs) are extended Ly$\alpha$ emitting nebulae, primarily found in galaxy over-dense regions at $z\sim1-3$ (e.g., \cite{steidel00,matsuda04,matsuda09,matsuda11,dey05,prescott08,barger12,valentino16,caminha16,cai17}).
It has been found that some 100 kpc-scale LABs are bright in submillimeter/millimeter wavelengths, suggesting a possible connection between LABs and dusty star-forming galaxies (DSFGs, e.g.,\cite{chapman01,chapman04,geach05,geach14,tamura13,yang14,hine16,alexander16,ao18,umehata17a,umehata17b,umehata18}).
However, we don't know if the DSFGs in LABs differ from typical DSFGs population.

Molecular gas depletion time ($\tau_{\rm{dep}}$) is given by the ratio of molecular gas mass ($M_{\rm{gas}}$) over star-formation rate (SFR), and is useful to investigate the evolutionary stage of DSFGs (e.g., \cite{chapman04,greve05,hodge13,ginolfi17}).
Typically DSFGs have a $\tau_{\rm{dep}}\lesssim10^8\,\rm{yr}$, while less active high-redshift star-forming galaxies have much longer timescales (e.g., \cite{tacconi17}).
Since there are very few LABs where molecular gas has been detected, it is unclear which evolutionary stage of dusty star-formation exists in LABs, and how long dusty star-formation of LABs will continue.
While LABs would be the seeds of galaxy groups and experience very rapid galaxy assembly (e.g., \cite{prescott12,kubo16,badescu17}), it is yet to be understood what they evolve into.

For the study of the interstellar medium (ISM), the dust emissivity index ($\beta$) at submillimeter wavelengths is derived from the Rayleigh-Jeans tail of infrared (IR) spectral energy distribution (SED), which reflects the dust chemical compositions \citep{galliano18}.
The IR SED is often approximated by a modified black body model of $S_{\rm{obs}}=(M_{\rm{d}}/D^2_{\rm{L}})\,\kappa_0(\nu/\nu_0)^{\beta}\,B_{\nu} (\nu_{\rm{rest}},T_{\rm{d}})$ \citep{hildebrand83}, where $M_{\rm{d}}$ is dust mass, $D^2_{\rm{L}}$ is luminosity distance, $\kappa_0$ is mass absorption coefficient at frequency $\nu_0$, $\beta$ is its variation as a function of frequency, and $B_{\nu} (\nu_{\rm{rest}},T_{\rm{d}})$ is the Planck function.
Different materials can produce different β, for instance, Crystalline Silicate have $\beta=2.0$ and Amorphous Graphite have $\beta=1.0$, which both change with temperature and frequency range (e.g., \cite{jones02,meny07}).

The dust emissivity index $\beta$ has also been studied in relation to galaxy properties in the local Universe.
\citet{boselli12} reports lower $\beta$ in low metallicity galaxies and \citet{cortese14} reports higher $\beta$ in local gas-poor, early-type galaxies with \textit{Herschel}/SPIRE.
While local galaxies have $\beta\sim$1.0-2.5 (e.g., \cite{dunne00,smith13,clements18}), it has been difficult to measure $\beta$ for high-redshift galaxies.
The high sensitivity of the Atacama Large Millimeter/submillimeter Array (ALMA) now allows us to constrain $\beta$ for high-redshift galaxies (e.g., \cite{tadaki17}).

Our target is SSA22 LAB18 at R.A. (J2000) = 22h17m29.0s, decl. (J2000) = +00$^\circ$07$'$50$''$ \citep{matsuda04} in the SSA22 protocluster at $z=3.1$ \citep{steidel98}.
LAB18 has a Ly$\alpha$ luminosity of $L_{\rm{Ly\alpha}}=(0.8 \pm 0.2) \times10^{43}$\,erg\, s$^{-1}$, physical size of 100 $\times$ 30\,kpc, and spectroscopic redshift of $z_{\rm{Ly\alpha}}=3.104$ \citep{matsuda11}.
LAB18 has been detected by using James Clerk Maxwell Telescope (JCMT) \citep{chapman01,chapman04,geach05,hine16,ao18} and \textit{Chandra} \citep{lehmer09,geach09}.
An X-ray detection with \textit{Chandra} observations suggests that LAB18 likely has a rapidly growing black hole in addition to the very rapid galaxy growth.
ALMA 850\,$\mu$m dust continuum observations (\cite{ao18}, Matsuda et al.\,in\,prep) identified four continuum sources (LAB18.a\,,b\,,c\,,d) toward the asymmetric, long filamentary structure of Ly$\alpha$ emission of LAB18 (Figure~\ref{fig1} left).
Since LAB18 is the brightest at 850\,$\mu$m among the LABs in the SSA22 protocluster, this source is the best target to conduct CO and multi-band dust continuum observations to investigate $\tau_{\rm{dep}}$ and $\beta$.
By combining these values, we can constrain the evolutionary stage of LABs and DSFGs.
We use the following cosmological parameters: $\Omega_m = 0.315,\ \Omega_\Lambda = 0.685,\ h = 0.67$ \citep{planck14a}.
In this cosmology, $1\farcs0$ corresponds to 7.9~kpc in physical length at $z = 3.1$.


\section{OBSERVATIONS}
\begin{table}
\caption{Galaxy properties for LAB18.b at $z=3.1$.}
\center
\scalebox{0.75}[0.75]{
\begin{tabular}{ccc}
\hline
 ID & Units &LAB18.b \\
 \hline
  RA & (J2000) & 22:17:28.94  \\
  Dec & (J2000) & +00:07:47.0 \\ 
 $z_{\rm{CO(4-3)}}$ & -- & $3.093\pm0.001$ \\
 $S_{2.7\,\rm{mm}}$ & ($\mu$Jy) & $52 \pm 11$ \\ 
 $S_{3\,\rm{mm}}$ & ($\mu$Jy) & $29 \pm 5$ \\
 $S_{\rm{CO(4-3)}}dv$ & (Jy\,km\,s$^{-1}$) & $0.083 \pm 0.006$ \\
 FWHM$_{\rm{CO(4-3)}}$ & (km\,s$^{-1}$) & $407 \pm 142$\\ 
 $L'_{\rm{CO(4-3)}}$  & (10$^9$\,K\,km\,s$^{-1}$\,pc$^2$) & $2.3 \pm 0.2$ \\
$M^a_{\rm{gas}}$ & ($10^9\,M_{\rm{\odot}}$) & $4.4^{+0.9}_{-0.6}$ \\
  $M^b_{\rm{dust, 3\,mm}}$ & $(10^7\,M_{\rm{\odot}}$) & $4.8 \pm 0.8$\\
 $\delta^c_{\rm{GDR}}$ & -- & $93^{+19}_{-14}$\\
 $T_{\rm{dust}}^d$ & (K) & $31.7\pm4.1$\\
 $\beta^d$ & (--) & $2.3\pm0.2$\\
 $L^d_{\rm{IR}}$ &  ($10^{12}\,L_{\rm{\odot}}$) & $2.7 \pm 1.6$ \\
 SFR$^d$ & ($M_{\rm{\odot}}$\,yr$^{-1}$) & $269 \pm 158$ \\
 $\tau_{\rm{dep}}$ & (Myr) & $17\pm10$ \\
\hline
\end{tabular}}
\footnotesize{
\begin{flushleft}
{\textbf{Notes}.\\
$^a$Gas mass with $r_{41}$=$0.41\pm0.07$ and $\alpha=0.8$\,$M_{\rm{\odot}}$ (K\,km\,s$^{-1}$\,pc$^2$)$^{-1}$.\\
$^b$Dust mass with $\beta=2.3$, $T_{\rm{d}}=31.7$\,K and $\kappa_{850}=3.2$\,cm$^{2}$\,g$^{-1}$.\\
$^c$Gas-to-dust mass ratio.\\
$^d$The errors come from the IR SED fitting.}
\end{flushleft}} 
\label{tab2}
\end{table}

We observed the LAB18 in ALMA Cycle 4 project (ID: 2016.1.01101.S; PI: Y. Kato).
The ALMA Band 3 observations were carried out through 12th to 14th November 2016 with 39--41 antennas with the baseline lengths of 15--1039\,m ($\sim$4--290\,k$\lambda$) in the dual-polarization setup.
The total on-source integration time was $\sim$5 hours.
The spectral windows were set to $\sim$97--101 GHz and $\sim$109-113 GHz, which covers CO\,$J=4-3$ line from $z_{\rm{CO}}=3.08-3.23$.
J2148+0657 and J0006--0623 were observed as flux calibrators.
The bandpass and phase were calibrated with J2148+0657 and J2226+0052, respectively.
The accuracy of absolute flux calibration is within 10\%.

We reduced the data with the Common Astronomy Software Applications ({\sc CASA}; \cite{mcmullin07}) 4.7.2 package in a standard manner.
From the calibrated data produced by the pipeline, we made primary beam corrected line free 3\,mm and 2.7\,mm continuum images with 97--101\,GHz and 109--113\,GHz bands, and a continuum subtracted spectral cube with a 80\,km\,s$^{-1}$ velocity resolution using {\sc clean} with natural weighting.
The synthesized beam full-width at half maximum (FWHM) of 3\,mm and 2.7\,mm continuum images are $1\farcs20\times1\farcs08$ with PA=65$^{\circ}$ and $1\farcs05\times0\farcs98$ with PA=67$^{\circ}$, respectively.
The achieved typical synthesized beam FWHM for CO spectral cube is $1\farcs07\times0\farcs99$ with PA=67$^{\circ}$ at 112.65 GHz.
The rms noises of the 3\,mm and 2.7\,mm continuum images are 5.9\,$\mu$Jy\,beam$^{-1}$ and 8.7\,$\mu$Jy\,beam$^{-1}$, respectively.
The pixel size is set to $0\farcs1$ for all data.

\section{RESULTS} 
\subsection{CO\,$J=4-3$ line}
The CO\,$J=4-3$ emission from LAB18.b is detected at 11\,$\sigma$ (peak flux to map variance on the integrated map). 
The derived systemic redshift is $z_{\rm{CO}}=3.093\pm0.001$ by using single Gaussian fitting (Figure~\ref{fig1} right).
We measured the total flux density of CO\,$J=4-3$ by using {\sc imfit} task of {\sc CASA} on the spectrally integrated flux map with primary beam correction.
We checked the total flux density of CO\,$J=4-3$ by using two components Gaussian fitting on the spectrum data, and found that the result does not significantly change.
Following \citet{solomon05}, we estimated the CO\,$J=4-3$ luminosity of $L'_{\rm{CO(4-3)}}=(2.3\ \pm\ 0.2)\times10^9$\,K\,km\,s$^{-1}$\,pc$^2$  (Table~\ref{tab2}).


\begin{figure}
\begin{center}
\includegraphics[scale=0.275]{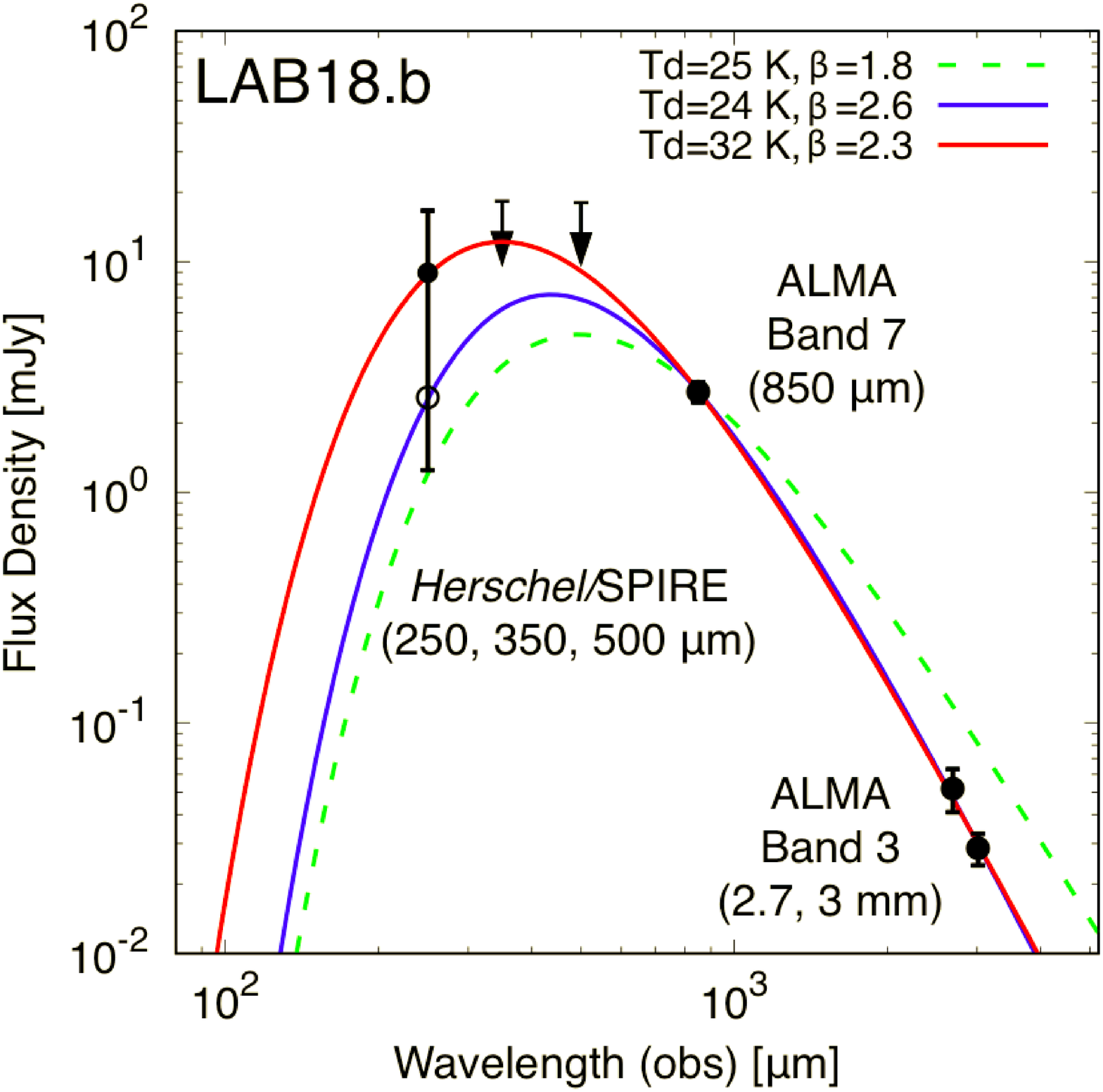}
\end{center}
\caption{The infrared (IR) spectral energy distribution (SED) for LAB18.b based on \textit{Herschel}/SPIRE and ALMA photometry.
We use a single temperature, optically thin modified black body model.
The best-fit values are $T_{\rm{d}}= 31.7 \pm 4.1$\,K, $\beta=2.3 \pm 0.2$, and $L_{\rm{IR}}=(2.7\pm1.6)\times10^{12}\,L_{\rm{\odot}}$ (red solid curve).
Dashed curve is for the supplementary purpose and the open circle at the SPIRE 250\,$\mu$m decreases by a factor of three for its filled circle (\S3.2).
}
\label{fig3}
\end{figure}

The CO\,$J=4-3$ luminosity is used to derive the molecular gas mass with $M_{\rm{gas}}=r_{41}\alpha_{\rm{CO}}L'_{\rm{CO(4-3)}}$, which gives a molecular gas mass of $M_{\rm{gas}}=(4.4^{+0.9}_{-0.6})\times10^{9}\,M_{\rm{\odot}}$ for LAB18.b (Table~\ref{tab2}).
We adopted luminous submillimetre galaxies (SMGs) median CO\,$J=4-3$ luminosity to CO $J=1-0$ luminosity ratio of $r_{41}=0.41\pm0.07$ \citep{bothwell13} and starbursts and mergers conversion factor of $\alpha_{\rm{CO}}=0.8$\,$M_{\rm{\odot}}$ (K\,km\,s$^{-1}$\,pc$^2$)$^{-1}$ \citep{downes98} since LAB18.b has a comparable IR luminosity.
We note that the molecular gas mass increases by a factor of five given the so-called Galactic conversion factor of $\alpha=4.36$\,$M_{\rm{\odot}}$ (K\,km\,s$^{-1}$\,pc$^2$)$^{-1}$.

The CO\,$J=4-3$ spectrum for LAB18.b shows double-peaked structure (Figure~\ref{fig1} right) and small spatial offset ($0\farcs4$ or $\sim$\,3\,kpc) between the red and blue components (Figure~\ref{fig1} middle).
These suggest that LAB18.b has a rotating disk or a merger, while the beam FWHM is significantly larger than the offset.

\subsection{Dust continuum}

\begin{figure*}
\begin{center}
\includegraphics[scale=0.4]{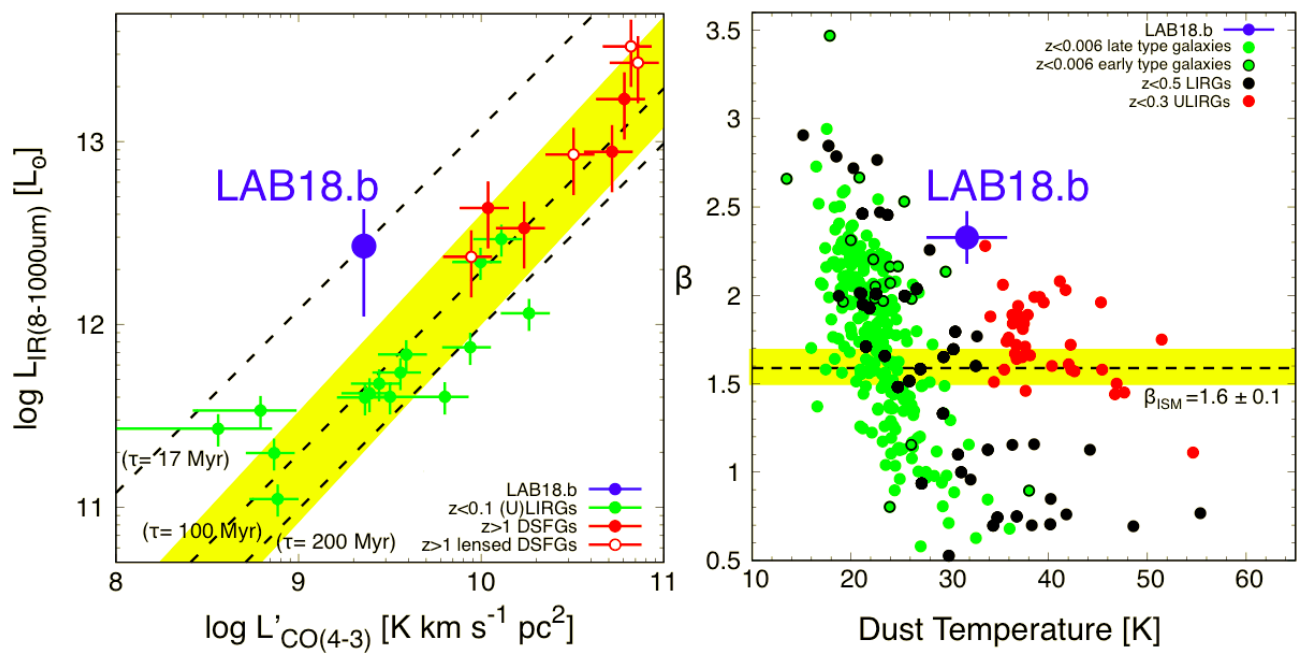}
\end{center}
\caption{\textit{(left)}.
$\it{L}_{\rm{IR}}$ vs. $L'_{\rm{CO(4-3)}}$ plot.  
The yellow shaded region is the best-fit relation and its scatter in \citet{greve14}.
The labeled dotted lines show constant molecular gas depletion times. 
LAB18.b shows one order of magnitude smaller  $\tau_{\rm{dep}}$ than typical DSFGs and (U)LIRGs in \citet{greve14}.
IR luminosities for $z>1$ lensed DSFGs are corrected for magnification (\cite{greve14}).
\textit{(right)}. $\beta$ vs. $T_{\rm{d}}$ plot.
LAB18.b has a high $\beta$ compared with the local ULIRGs (red points; \cite{clements18}) and the Galactic value (yellow shaded region of $\beta=1.6\pm0.1$; \cite{planck14b}).
The location of LAB18.b is between local ULIRGs and local gas-poor, early-type galaxies (green points with black circles; \cite{cortese14}).
Black points (LIRGs) and green points (late type galaxies with $L_{\rm{IR}}=10^8-10^{11}\,L_{\rm{\odot}}$) are compiled from \citet{smith13} and \citet{cortese14}, respectively.
All data points from literatures use \textit{Hershcel}/SPIRE bands for IR SED fitting with isothermal modified black body model.}
\label{fig4}
\end{figure*}

We detected 2.7\,mm and 3\,mm dust continuum emission at the position of the LAB18.b (Figure~\ref{fig1} left).
We measured the total flux density of both dust continuum emission using {\sc imfit} task of {\sc CASA} with primary beam correction (Table~\ref{tab2}). 
We derived an infrared (IR) luminosity $\it{L}_{\rm{IR\,[8-1000\,\mu m]}}$, a dust temperature $T_{\rm{d}}$ and a dust emissivity index $\beta$ by fitting a single temperature, optically thin modified black body model to the SPIRE 250\,$\mu$m \citep{kato16}, ALMA Band 7 850\,$\mu$m (\cite{ao18}; Matsuda et al.\,in\,prep), ALMA Band 3 2.7\,mm, and 3\,mm dust continuum emission (Figure~\ref{fig3}).
The SPIRE 250\,$\mu$m flux density is measured at the position of LAB18.b and applied a source confusion noise for photometry error \citep{kato16}.
The accuracy of the ALMA Band 7 850\,$\mu$m absolute flux calibration is evaluated to be within 10\%, which is applied for the SED fitting (Figure~\ref{fig3}).
The best-fit values are $T_{\rm{d}}= 31.7 \pm 4.1$\,K, $\beta=2.3 \pm 0.2$, and $L_{\rm{IR}}=(2.7\pm1.6)\times10^{12}\,L_{\rm{\odot}}$.
These errors show 68\% confidence intervals of the $\chi^2$ SED fits to the photometry data.
We derived a star-formation rate of SFR$=269\pm158$\,$M_{\rm{\odot}}$\,yr$^{-1}$ from the empirical $\it{L}_{\rm{IR}}$-SFR relation in \citet{kennicutt98} and Chabrier IMF \citep{chabrier03}; SFR=$1.0\times10^{-10} \it{L}_{\rm{IR}}$, where the units of SFR and $\it{L}_{\rm{IR}}$ is $M_{\rm{\odot}}$\,yr$^{-1}$ and $L_{\rm{\odot}}$, respectively.
The dust mass is also estimated with a relation of $M_{\rm{d}} =S_{\rm{obs}}\,D_L^{\rm{2}}/(\kappa_{\rm{d}}(\nu_{\rm{rest}})\,B_{\nu} (\nu_{\rm{rest}},T_{\rm{d}})\,(1+z))$; the mass absorption coefficient is $\kappa_{\rm{d}}(\nu_{\rm{rest}})=\kappa_{850}(\nu/\nu_{850})^{\beta}$, where $\kappa_{850}$ is assumed to be $\kappa_{850}=3.2$\,cm$^{2}$\,g$^{-1}$ \citep{demyk17}.
We derived the dust mass of $M_{\rm{d}}=(4.8\pm0.8)\times10^{7}\,M_{\rm{\odot}}$ using $S_{3\,\rm{mm}}$, which indicates gas-to-dust mass ratio of $\delta_{\rm{GDR}}=93^{+19}_{-14}$ (Table~\ref{tab2}).

We note the SPIRE 250\,$\mu$m photometry could be overestimated because of source blending with other ALMA 850\,$\mu$m sources.
If the SPIRE 250\,$\mu$m flux of LAB18.b is proportional to the ALMA 850\,$\mu$m flux as reported in \citet{ao18}, the flux decreases by a factor of three and the best-fit results are $T_{\rm{d}}=24.1\pm1.3$\,K, $\beta=2.6\pm0.1$ and $L_{\rm{IR}}=(1.3\pm0.3)\times10^{12}\,L_{\rm{\odot}}$ (Figure~\ref{fig3} open circle).
SPIRE source confusion does not change our results that LAB18.b has a high $\beta$ and $L_{\rm{IR}}$ at a given $L'_{\rm{CO(4-3)}}$ (Figure~\ref{fig4} left).

\section{DISCUSSION AND SUMMARY}
Is LAB18.b a typical DSFG at high-redshift?
The left panel of Figure~\ref{fig4} shows that LAB18.b has a low $L'_{\rm{CO(4-3)}}/L_{\rm{IR}}$ ratio among $z>1$ DSFGs and $z<0.1$ (Ultra) Luminous Infrared Galaxies (U)LIRGs in \citet{greve14}.
The labeled dotted lines show constant gas depletion times ($\tau_{\rm{dep}}=M_{\rm{gas}}/\rm{SFR}$) derived with  $r_{41}=0.41$, $\alpha=0.8$\,$M_{\rm{\odot}}$ (K\,km\,s$^{-1}$\,pc$^2$)$^{-1}$, and SFR conversion.
The derived molecular gas depletion time of LAB18.b is $\tau_{\rm{dep}}=17\pm10$\,Myr, which is one order of magnitude smaller than the typical values of $\tau_{\rm{dep}}=100-200$\,Myr for $z<0.1$ (U)LIRGs and $z>1$ DSFGs \citep{greve14}.
The right panel of Figure~\ref{fig4} shows that LAB18.b is located between local ULIRGs \citep{clements18} and local gas-poor, early-type galaxies \citep{cortese14}.
These suggest that LAB18.b is not a typical DSFG but in a transition phase from a DSFGs to a gas-poor, early-type galaxy.
It would be interesting to test if LABs are associated with DSFGs with short $\tau_{\rm{dep}}$ and high $\beta$ by future ALMA observations.

As shown in the right panel of Figure~\ref{fig4}, the $\beta$ of LAB18.b is larger than the typical values of ULIRGs \citep{clements18} and our Galaxy \citep{planck14b}.
What can produce such a high $\beta$ value?
It is known that β should be independent of the dust grain size since the observed IR wavelength is much larger than the typical dust grain size in interstellar medium (0.3\,nm $\lesssim r\lesssim$ 0.3\,$\mu$m; \cite{galliano18}).
However, a high $\beta$ can be produced by the chemical composition (e.g., \cite{miyake93}).
For instance, \citet{demyk17} showed that Mg-rich amorphous silicates reproduce $\beta>2.0$ and a large $\kappa_{850}\sim3.2$\,cm$^{2}$\,g$^{-1}$ compared with the typically assumed values of $\kappa_{850}=0.4-1.5$\,cm$^{2}$\,g$^{-1}$ in the diffuse interstellar medium (ISM) in the Galaxy and local galaxies \citep{dunne03}.
We estimated the $M_{\rm{d}}$ and $\delta_{\rm{GDR}}$ for LAB18.b by assuming this $\kappa_{850}=3.2$\,cm$^{2}$\,g$^{-1}$ (\S3.2 and Table~\ref{tab2}).  
We note that if we adopt $\kappa_{850}=1.0$\,cm$^{2}$\,g$^{-1}$, the dust mass increases by a factor of three, and results in a small gas-to-dust mass ratio of $\delta_{\rm{GDR}}\approx30$, which is much lower than the values in both local (U)LIRGs ($\delta_{\rm{GDR}}\approx120$; \cite{wilson08}) and distant submillimeter galaxies (SMGs) ($\delta_{\rm{GDR}}\approx90$; \cite{swinbank14}).
Thus, if LAB18.b has the typical $\delta_{\rm{GDR}}$ as local (U)LIRGs and SMGs, the large $\kappa_{850}=3.2$\,cm$^{2}$\,g$^{-1}$ expected from the high $\beta$ is reasonable.
We also note that if we estimate the dust mass with $S_{850\,\mu\rm{m}}$, standard $\kappa_{850}=1.0$\,cm$^{2}$\,g$^{-1}$ and $\beta=2.3$, the dust mass is still three times smaller than the typical DSFGs estimate (e.g., $S_{850\,\mu\rm{m}}$, standard $\kappa_{850}=1.0$\,cm$^{2}$\,g$^{-1}$ and $\beta=1.5$).
This would support that LAB18.b still ends up with a low dust mass even in a standard $\kappa_{850}$, and would support that LAB18.b has low gas mass if the gas-to-dust mass ratio is same as typical DSFGs.

We found that LAB18 has short $\tau_{\rm{dep}}$ and high $\beta$.
This suggests that DSFGs in LAB18 are transition phase to evolve gas-poor, early-type galaxies.
We argue that ALMA CO and multi-band dust continuum observations can constrain the evolutionary stage of high-redshift galaxies through $\tau_{\rm{dep}}$ and $\beta$.
The precise measurement of $\beta$ can also constrain the chemical composition and $\kappa$ of dust grains even in the early Universe, which is important to reliable estimate of dust mass.

\begin{ack}
We would like to thank Maciej Koprowski for careful reading our manuscript and for giving useful comments. 
We thank Ian Smail, Yoichi Tamura, Takaya Nozawa, and Akimasa Kataoka for the useful discussion.
We thank L. Cortese, D. J. B. Smith, and T. R. Greve for sharing their data.
Y.M, H.U, and Y.T acknowledge JSPS KAKENHI Grant Number 17H04831, 17KK0098, 17K14252 and 16H02166.
D.M.A acknowledges the Science and Technology Facilities Council (STFC) through grant ST/P000541/1.
Y.A. acknowledges partial support by NSFC grant 1373007.
M.H. acknowledges the support of the Swedish Research Council, Vetenskapsr{\aa}det and the Swedish National Space Board (SNSB), and is Fellow of the Knut and Alice Wallenberg Foundation.
This paper makes use of the following ALMA data: ADS/JAO.ALMA\#2016,1.01101.S, ADS/JAO.ALMA\#2013.1.00704.S.
ALMA is a partnership of ESO (representing its member states), NSF (USA) and NINS (Japan), together with NRC (Canada) and NSC and ASIAA (Taiwan) and KASI (Republic of Korea), in cooperation with the Republic of Chile.
The Joint ALMA Observatory is operated by ESO, AUI/NRAO and NAOJ.
This work was supported by the ALMA Japan Research Grant of NAOJ Chile Observatory, NAOJ-ALMA-0086.
\end{ack}


\end{document}